\documentclass[aps,prd,twocolumn,showpacs,groupedaddress]{revtex4}

\usepackage{graphicx}
\usepackage{color}

\begin{document}


\title{Time Dependent Quark Masses and Big Bang Nucleosynthesis Revisited}


\author{Myung-Ki Cheoun,$^{1}$\footnote{cheoun@ssu.ac.kr} Toshitaka
Kajino,$^{2,3,4}$ Motohiko Kusakabe$^{5}$ and Grant J. Mathews$^{6}$}

\affiliation{
$^1$Department of Physics, Soongsil University, Seoul 156-743, Korea\\
$^2$Division of Theoretical Astronomy, National Astronomical Observatory
of Japan, Mitaka, Tokyo 181-8588, Japan \\
$^3$Department of Astronomy, Graduate School of Science, University of
Tokyo,  Hongo, Bunkyo-ku, Tokyo 113-0033, Japan \\
$^4$Department of Astronomical Science, The Graduate University for
Advanced Studies, Mitaka, Tokyo 181-8588, Japan \\
$^5$Institute for Cosmic Ray Research, University of Tokyo, Kashiwa,
Chiba 277-8582, Japan\\
$^6$Department of Physics and Center for Astrophysics, University of
Notre Dame, Notre Dame, Indiana 46556, USA \\
}


\date{\today}

\begin{abstract}

We reinvestigate the constraints from primordial nucleosynthesis  on a possible  time-dependent quark mass.  The limits on such quark-mass variations are particularly sensitive to the adopted observational abundance constraints.  Hence, in the present study we have considered updated light-element abundances and uncertainties deduced from observations. We also consider  new nuclear reaction rates and an independent analysis of 
the influence of such quark-mass variations on the resonance properties of the important  $^3$He($d,p)^4$He reaction.
We find that the updated abundance and resonance constraints  imply a narrower  range on the possible quark-mass variations in the early universe.  We also find that, contrary to previous investigations, the optimum concordance region reduces to a (95\% C.L.) value of
$-0.005 \lesssim \delta m_q/m_q \lesssim 0.007$ consistent with no variation in the averaged quark mass.

\end{abstract}



\maketitle

\section{Introduction}

Big bang nucleosynthesis (BBN) remains as one of the key constraints on the physics of the early universe.  BBN occurred during  the epoch from $\sim 1-10^4$ sec into the big bang, and as such, is the only direct probe of new physics during the important radiation dominated epoch.  Moreover, once the nuclear reaction rates have been specified, the light-element abundances produced during the
big bang only depend \cite{BBNrev} upon a single parameter, the baryon-to-photon ratio $\eta$, which is now constrained to high precision from the effects of baryons on the second and third acoustic peaks in the power spectrum of cosmic microwave background fluctuations as deduced from data obtained with  the Wilkinson Microwave Anisotropy Probe ({\it WMAP}) \cite{WMAP}.   

 In this context there has been considerable interest \cite{Coc07,Dent07,Flambaum07,Flambaum08,Flambaum09,Berengut10} in recent years in the use of BBN to constrain  any possible  time variation of fundamental physical constants in the early universe.  One can only detect variations of dimensionless quantities which are independent of the units employed.  Therefore, of particular interest for the present work is the variation in the dimensionless ratio  $X_q \equiv m_q/\Lambda_{QCD}$, where $m_q \equiv (m_u + m_d)/2$ is the averaged quark mass and $\Lambda_{QCD}$ is the scale of quantum chromodynamics (defined \cite{Flambaum07} as the energy associated with the Landau
pole in the logarithm of the running strong coupling constant).  As in \cite{Flambaum07} we note that nuclear parameters
(e.g.~nucleon mass, reaction cross sections, etc.)  constrain $\Lambda_{QCD}$.  Hence, it is convenient
to assume that $\Lambda_{QCD}$ is constant and calculate the time variation dependence in terms of $\delta m_q / m_q \equiv \delta X_q / X_q$. $\Lambda_{QCD}$ can then be added at the end to reconstruct a dimensionless parameter.  
 
 A time dependence of fundamental constants in an expanding universe can be a generic result \cite{Flambaum08,Uzan03} of theories which attempt to unify gravity and other interactions.  Such grand unified theories imply  correlations among variations in all of the fundamental constants \cite{Flambaum08}.  However, it can be argued \cite{Flambaum07} that BBN is much more sensitive to variations in $X_q$ than other physical parameters such as variations in the fine structure constant.   Hence, $X_q$  may be the best parameter with which to search for evidence of time variation of fundamental constants in the early universe.  It is of particular note that a recent study \cite{Berengut10} found that an increase in the average quark mass by an amount  $\delta m_q/m_q = 0.016 \pm 0.005$  provides a better agreement between observed light-element primordial abundances than those predicted by the standard big bang.  A similar conclusion was reached in \cite{Dent07}.  Because of the importance of such evidence for a changing quark mass, it is important to carefully reexamine all aspects of the physics which has been used to place constraints upon this parameter from BBN.  That is the purpose of the present work.

   In addition to the interest in evidence for a variation of the physical constants in the early universe, it has also been suggested   \cite{Flambaum07,Berengut10} that such variations may provide insight into a fundamental problem in BBN.  There is an apparent discrepancy between the observed primordial abundance of $^7$Li and that inferred from BBN when the limits on the baryon-to-photon ratio $\eta$ from the WMAP analysis are adopted.  Indeed, standard BBN with the WMAP value for $\eta$ implies an abundance of $^7$Li that is a factor of $2.4-4.3$ (4-5 $\sigma$ C.L.) times higher \cite{Cyburt08} than the observationally deduced primordial abundance.  A great many possible solutions to this lithium problem have been proposed, (e.g.~\cite{Kusakabe10} and references therein).  Nevertheless, in view of the importance of the possibility that depleted $^7$Li indicated a variation of fundamental parameters, it is worth carefully reinvestigating this hypothesis.

In this paper, therefore, we carry out an independent evaluation of the effects on BBN from  
a variation in the parameter $\delta X_q/X_q$.  In this effort we are motivated by new detailed analyses \cite{BBNrev,Aver10} of the uncertainties in the observed light-element abundance constraints.  We also make an independent evaluation of the resonant  $^3$He($d,p$)$^4$He reaction rate based upon both the forward and reverse reaction dependence on $\delta m_q / m_q$.  We find that, although the uncertainty in the results increases due to variations in the resonance parameters,  the revised abundance constraints narrow the range of possible variations in the quark mass from  BBN.  This latter constraint dominates and  decreases the optimum concordance region to a value of $-0.005 \lesssim \delta m_q/m_q \lesssim 0.007$.  Although some variation in the quark mass is not ruled out, the  results of the present study are  consistent with no variation in the averaged quark mass.

\section{Model}\label{sec2s}
 The calculations performed here are based upon the standard 
BBN network code of \cite{kawano92,Smith93}, with a number of 
reactions for light nuclei ($A \le  10$) updated based upon 
the latest JINA REACLIB Database V1.0 \cite{Cyburt10}, and/or the NACRE rates \cite{Descouvemont:2004cw}.  The rate of the 
$^4$He($^3$He,$\gamma$)$^7$Be reaction was additionally replaced with that of
\cite{Cyburt:2008up}. We also adopt the  neutron
lifetime of $878.5 \pm 0.7_{stat} \pm 0.3_{sys}$ from \cite{Serebrov10}.  The usually adopted world average value of $885.7 \pm 0.8~s$ is based upon an average deduced by the Particle Data Group \cite{PDG}  in 2006 and repeated in 2010.  In recent work \cite{Serebrov10}, however, it has been demonstrated via Monte Carlo simulations of the experiments that some of the data used in that average suffer from a $- 6~s$ systematic error.  Moreover, the lower neutron lifetime is based upon a number of improvements  \cite{Serebrov05} in the measurement.   This shorter lifetime also better satisfies the  unitarity test
 of the Cabibbo-Kobayashi-Maskawa matrix \cite{Serebrov05}, and it improves the general agreement between observed primordial abundances and  BBN \cite{Mathews04}.

As described below, we make use of an analytic scaling of the resonant $^3$He($d,p$)$^4$He reaction rate from \cite{Cyburt:2004cq} because it  explicitly includes the relevant $Q$ value and resonance energy dependences.  We have modified  the thermonuclear
reaction rates based upon the procedure outlined in \cite{Dent07,Berengut10} with a refinement for resonant reactions as described below.
 In particular, we make extensive use of the relation between nuclear parameters and the standard model of particle physics as delineated  in \cite{Dent07}.  
  
 \subsection{Binding Energy Dependence}
 Changing $X_q$ affects nuclear reaction rates in several ways.  The strongest influence on BBN abundances, however, simply comes from  the dependence of reaction rates on nuclear binding energies.
 The main effect of nuclear binding energies enters through the sensitive dependence of reaction rates on the reaction $Q$ values in the exit channel.  This is because the $Q$ value affects the energy of the outgoing products  for  reactions with a positive $Q$ value.  Whenever possible  in the calculations described below, we modify the $Q$-value directly, $Q \rightarrow Q + \delta Q$ in applicable  formulas for the reaction rates.  Nevertheless, for some rates, the $Q$-value dependence is not apparent and it is necessary  to 
specify explicitly the effects of varying  the $Q$ value for the relevant reactions.

  The $Q$-value effects on the reaction rates are qualitatively described in \cite{Berengut10}, which we now summarize and slightly generalize.  For  electric dipole radiative capture  the cross section for photons of polarization $m$ and energy $E_\gamma$ can be written \cite{Christy61} as
  \begin{equation}
  \sigma_{1m} = \frac{16 \pi}{9} \biggl( \frac{E_\gamma}{\hbar c}\biggr)^3 \frac{1}{\hbar v} \vert Q_{1 m}\vert^2 ~~,
  \end{equation}
  where $\vert Q_{1 m}\vert$ is the electric dipole moment and $v$ is the relative velocity in the entrance channel.
  At low energies  the dominant $Q$-value dependence  simply derives from the exit channel  $E_\gamma^3$ term \cite{Dent07}.  Hence,  one can write
\begin{equation}
\label{eq:E1_capture}
\sigma (E) \propto E_\gamma^3 \sim (Q+E)^3 ~,
\end{equation}
where $E$ is the reaction center of mass energy and $E_\gamma$ is the energy of the emitted gamma ray.  

For low-energy reactions in which there are  two nuclear species in the final state, the reaction cross section depends upon the resonance penetrabilities in both the entrance channel (for resonant reactions) and the exit-channels \cite{FCZ67}.   In the exit-channel the $Q$-value dependence arises from two sources.  For one, the cross section becomes proportional to the final state velocity, $v \propto (Q+E)^{1/2}$ of the two emitted particles.  For the other, one must also take into account the penetrabilities of the two exit-channel particles if they both carry charge.  In this case the penetrability also  depends upon the reaction $Q$ value.   Hence, one can write \cite{Dent07}
\begin{equation}
\sigma (E)\sim (Q+E)^{1/2} e^{-\sqrt{E_G/(Q+E)}}~~.
\label{eq:sigma_transfer}
\end{equation}
where  $E_{G}$ is the Gamow energy characterizing the exit-channel penetrability, 
\begin{equation}
E_{G} =\frac{2\pi^2 e^4}{\hbar^2}W^2~~, 
\label{eq:Gamov Energy}
\end{equation}
with
\begin{equation}
W = Z_{3}^2Z_{4}^2 \mu ~~,
\end{equation}
while $Z_3$ and $Z_4$ are the exit-channel charges.

At the  temperatures associated with BBN ($kT < 100$ keV), one generally has $E \ll Q$. Therefore, one can expand Eq.~(\ref{eq:sigma_transfer}) in $Q$ to find the dependence of the  cross section on changes in $\delta Q$,
\begin{equation}
\label{eq:sigma_transfer_linear}
\sigma = \sigma_0 \left[ 1 + \frac{1}{2}\left( 1+ \sqrt{\frac{E_G}{Q}}\right)
		    \frac{\delta Q}{Q} + ... \right]~~.
\end{equation}
Although the Gamow term in Eq.~(\ref{eq:sigma_transfer}) is usually  small, it can be important for some reactions,  e.g. $^7$Be($n,p$)$^7$Li or $^3$He($n,p$)$^3$H \cite{Berengut10}.

An additional sensitivity on the reaction $Q$ values occurs for the
reverse photodisintegration rates which are related to the forward rates via the equation of  detailed balance:  
\begin{equation}
{\langle \sigma v \rangle_{\rm rev}} \propto  {\langle \sigma v
 \rangle_{\rm fwd}}
\times  e^{{-Q}/{kT}}~~.
\end{equation}

For the special case of the  $p(n,\gamma)$D reaction, there is also an additional sensitivity to the position of a virtual level with energy $\epsilon_\nu = 0.07$~MeV. This leads to  \cite{Dmitriev04}
\begin{equation}
\label{eq:npdg}
\left< \sigma v \right> \sim \left[ 1 + \left( 5/2 + \sqrt{\frac{Q}{\epsilon_\nu}}\right) \frac{\delta Q}{Q} \right]\ .
\end{equation}

As in \cite{Berengut10}, we denote the sensitivity of nuclear binding energies to the averaged light-current quark mass $m_q$ by
\begin{equation}
\label{eq:K}
K = \frac{\delta E / E}{\delta m_q / m_q}\ .
\end{equation}
 The $K$ values for light nuclei have been summarized   in \cite{Dent07,Flambaum07,Flambaum09}. As in \cite{Berengut10} we use the values given by the AV18+UIX nuclear two-body interaction, with hadron mass variations calculated in terms of the $m_q$ using the Dyson-Schwinger equation calculation of \cite{Flambaum06}.

\subsection{Effect on Nuclear Resonances}\label{sec2_2s}

We will treat the resonance reactions of BBN slightly different than that of  \cite{Berengut10}.   There are two reactions in BBN which are dominated by narrow resonances.  These are the $^3$He($d,p$)$^4$He and $^3$H($d,n$)$^4$He reactions. The cross sections for these reactions can be  factored into the usual separation of a penetrability and the astrophysical $S$ factor:
\begin{equation}
\sigma \left(E\right)=S\left(E\right)\frac{1}{E}e^{-\sqrt{\frac{E_{G}}{E}}},
\label{eq:Cross Section}
\end{equation}
where $S\left(E\right)$ contains information on the nuclear matrix element. For a resonant reaction, the $S$-factor can be written 
\begin{equation}
S(E) = \frac{P(E)}{(E-E_r)^2 + \Gamma_r^2/4}~~,
\end{equation}
where  $E_r$ is the resonance energy and $\Gamma_r$ is the total
resonance width.  Any  residual energy dependence beyond that of the
entrance-channel resonance (e.g. the direct capture contribution, and the penetrability in the exit channel) is usually contained in a
polynomial  $P(E)$ that is  fit to the measured  cross section. We adopt
the unperturbed cross-section parameters  for these reactions in \cite{Cyburt10,Cyburt:2004cq}.  This  gives resonance properties of  $E_r = 0.183$~MeV and  $\Gamma_r = 0.256$~ MeV for the $^3$He($d,p$)$^4$He reaction, while  $E_r = 0.0482$~MeV and  $\Gamma_r = 0.0806$~MeV for the $^3$H($d,n$)$^4$He reaction.  The ground-state binding energies were taken from the experimental data in \cite{Flambaum07}.   

Changes to the $^3$He($d,p$)$^4$He reaction will affect the abundances of  the primordial abundances of $^3$He and $^7$Be. Changes in the $^3$H($d,n$)$^4$He reaction similarly  affect the abundances of $^3$H and $^7$Li.   There are, however, newer formulations for these resonant rates given in \cite{Cyburt10} and \cite{Descouvemont:2004cw}.  These revised  rates incorporate newer data and an improved R-matrix fit to the data.  However, these newer rates are given in an analytic form that does not explicitly manifest the total dependence on reaction $Q$ value and resonance energy $E_r$.  To account for this, we adopt new rates but correct them using a scaling based upon the discussion above and the   analytic reaction rate given in \cite{Cyburt:2004cq}.

Explicitly, in \cite{Cyburt:2004cq}, the thermonuclear reaction rate for a resonant reaction can be written
\begin{eqnarray}
 N_A \langle \sigma v \rangle = \frac{C \sigma_0}{(kT)^{3/2}} \exp{\biggl[ - \biggl(\frac{27 E_G}{4 kT}\biggr)^{1/3}}\biggr] \nonumber \\
\times  \frac{S_{eff}(E_0)}{1 + [(E_0 - E_r)/(\Gamma_r/2)]^2}~~,
\label{cyburtsigv}
\end{eqnarray}
where $C = N_A (8/\mu \pi)^{1/2}$ is a normalization.  For the reactions with a two-body charged final state of interest here, the dependence on the exit-channel penetrability [Eq.~(\ref{eq:sigma_transfer})] is contained in a polynomial expansion of $S(E)$ in $E$.  The peak of the Gamow window in the Gaussian approximation is $E_0 = E_G^{1/3} (kT/2)^{2/3}$, so that the penetrability in the exit channel is contained in a polynomial expansion of $S_{eff}(E_0)$.  Hence, combining Eqs. (\ref{eq:sigma_transfer_linear}) and (\ref{cyburtsigv}), a general scaling for any resonant reactions with a charged two-body final  state can be written
\begin{eqnarray}
\bigl[ N_A \langle \sigma v \rangle \bigr]  &=& \bigl[ N_A \langle \sigma v \rangle \bigr]_0 \times \left[ 1 + \frac{1}{2}\left( 1+ \sqrt{\frac{E_G}{Q}} \right) 
		    \frac{\delta Q}{Q} + ... \right] \nonumber \\
		     &&\times\biggl(\frac{1 + [(E_0-E_r^0)/(\Gamma_r/2)]^2}{1 + [(E_0-E_r)/(\Gamma_r/2)]^2}\biggr)~~,
\end{eqnarray}
where $[ N_A \langle \sigma v \rangle]_0$ is the unperturbed reaction, and $E_r^0$ is the unperturbed resonance energy.

\begin{figure}
\begin{center}
\includegraphics[width=8.0cm,clip]{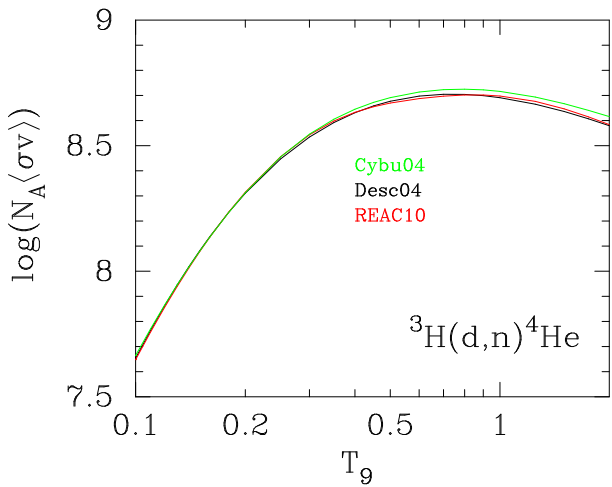}
\end{center}
\caption{Unperturbed ($\delta m_q/m_q= 0)$ reaction rates $[ N_A \langle \sigma v \rangle]_0$ for the $^3$H($d,n$)$^4$He reaction as given in \cite{Cyburt:2004cq} (Cybu04),  \cite{Cyburt10} (REAC10), and \cite{Descouvemont:2004cw} (Desc04) as a function of temperature $T_9\equiv T/(10^9~{\rm K})$.
}\label{fig:1}
\end{figure}

  The analytic forms for the unperturbed reaction rates given in   \cite{Cyburt10}, \cite{Descouvemont:2004cw}  and \cite{Cyburt:2004cq} appear to be quite  different.
 Nevertheless, in spite of the improvements  in experimental data and theoretical fit adopted, the unperturbed rates are surprisingly similar.  
 Figure \ref{fig:1} shows a comparison of the three different unperturbed resonant rates  over the range of temperatures of $T_9\equiv T/(10^9~{\rm K}) = 0.1 $ to $2$.  Although there is some slight difference in these rates  at high temperature ($T_9 \ge 0.5$), in the temperature range of most relevance to  BBN ($T_9 \lesssim 1$), the three rates are essentially indistinguishable.  Hence, in this study we have confirmed that there is no significant uncertainty in the constraints on $\delta m_q/m_q$ introduced by the more recently deduced formulations \cite{Cyburt10} and \cite{Descouvemont:2004cw} for the  resonant reaction  rates.

As noted above, the dependence of the resonance energy $E_r$ on variations  of $\delta m_q$ have been evaluated in Refs.~\cite{Dent07,Berengut10} and are adopted in the present study.  To determine the sensitivity of the BBN resonant reactions to changes in $\delta m_q$, one begins by expanding the resonance energy in terms of the unperturbed and perturbed energies, i.e.,
\begin{equation}
 E_r \rightarrow E_r^0 + \delta E_r~~.
\label{eq:Er0} 
\end{equation}
 For the $^3$He($d,p$)$^4$He reaction, the resonance is an excited state of the compound nucleus $^5$Li$^*$.
 For the $^3$H($d,n$)$^4$He reaction the compound resonance is $^5$He$^*$.
 The observed unperturbed resonance excitation energies for Eq.~(\ref{eq:Er0}) are  $E^0_{^5{\rm
Li}^{*}} = -9.76$~MeV and $E^0_{^5{\rm He}^{*}} = -10.66$~MeV. 

The resonance energies are related to the excitation energy in the compound nucleus and the net binding energies of the reactants, i.e.,
\begin{eqnarray}
\label{eq:Er1}
E_r^{(d,p)} &=& E_{^5{\rm Li}^{*}} - E_{^3{\rm He}} - E_d \\
\label{eq:Er2}
E_r^{(d,n)} &=& E_{^5{\rm He}^{*}} - E_t - E_d~~.
\end{eqnarray}
The net shift  in the resonance position due to variations in $m_q$ is then the combination of the shift of three energies,
\begin{eqnarray}
\delta E_r^{(d,p)} &=& \delta E_{^5{\rm Li}^{*}} - \delta E_{^3{\rm He}} - \delta E_d \\
 =&& \left( K_{^5{\rm Li}^{*}} E_{^5{\rm Li}^{*}} - K_{^3{\rm He}}
      E_{^3{\rm He}} - K_d E_d \right)
     \frac{\delta m_q}{m_q}~~. \nonumber
\end{eqnarray}
with the $K$ values defined by Eq.~(\ref{eq:K}) and summarized in \cite{Dent07}.

Here we point out that there is a consistency check on the $\delta m_q$ sensitivity of the forward $^3$He($d,p$)$^4$He reaction from the reverse $^4$He($p,d$)$^3$He reaction.  For this case we have
\begin{eqnarray}
\delta E_r^{(p,d)} &=& \delta E_{^5{\rm Li}^{*}} - \delta E_{^4{\rm He}} - \delta E_p \\
 &=& \left( K_{^5{\rm Li}^{*}}E_{^5{\rm Li}^{*}}   - K_{^4{\rm
      He}}E_{^4{\rm He}} - K_p E_p \right) \frac{\delta m_q}{m_q}~~. \nonumber
\end{eqnarray}
 As a test on the robustness of the constraint on $\delta m_q/m_q$ we
 also include the variations in the resonance energy based upon the parameters in  this reverse channel.  For this case we
 have $K_{^5{\rm Li}^*}=-3.131$, 
$K_{^5{\rm He}^*}=-2.867$, and the constraint: $
\delta E_r^{(N,d)}=\delta E_{^5A^*} -\delta E_{^4{\rm He}} =0$ for $N=n$ or $p$. This implies that 
$K_{^5A^*}=K_{^4{\rm He}}\times(E_{^4{\rm He}}/E_{^5A^*})=-1.08\times (-28.30)/E_{^5A^*}$.

  \subsection{Primordial Light-Element Abundance Constraints}
  Before summarizing the results of our BBN calculation and the constraints on $\delta m_q/{m_q}$,  it is useful to first summarize the 
light-element abundance constraints.  In this task we are much aided by a recent thorough review by Iocco et al.~\cite{BBNrev}, and also new constraints on the primordial helium abundance \cite{Aver10}.  Our adopted abundance constraints are then as follows.

\subsubsection{Deuterium}
Deuterium is best measured in the spectra of narrowline
Lyman-$\alpha$ absorption systems in the foreground of high redshift Quasars.  Unfortunately, only 14 such systems have been found \cite{BBNrev}.  Taken altogether  they exhibit an unexpectedly large dispersion.  This suggests that there could be unaccounted systematic errors.  This enhanced error can be approximately accounted for by constructing the weighted mean and standard deviation directly from the data points.  Based upon this, we adopt the conservative range for the primordial deuterium abundance of ${\rm D/H} = (2.87^{+0.22}_{-0.19})\times 10^{-5}$ from \cite{BBNrev}.  This implies a  $2 \sigma$ (95\% C.L.)  concordance region of
\begin{equation}
2.49 \times10^{-5}< {\rm D}/{\rm H}< 3.31\times10^{-5}~~.
 \label{eq21s}
\end{equation}

  We note, however, that if one restricts the data to the six well resolved systems for which there are multiple Lyman-$\alpha$ lines  \cite{Pettini08}, one slightly lowers the $2 \sigma$  deuterium constraint to
\begin{equation}
2.44 \times10^{-5}< {\rm D}/{\rm H}< 3.22\times10^{-5}~~.
 \label{eq21t}
\end{equation}
Adopting this lower constraint would strengthen the conclusions of this paper.  Nevertheless, we adopt the more conservative constraint of Eq.~(\ref{eq21s}) based \cite{BBNrev} upon all of the data.

\subsubsection{$^3$He}
The abundance of $^3$He is best measured in Galactic HII regions by the 8.665~GHz
hyperfine transition of $^3$He$^+$~\cite{Bania02}.  A plateau
with a relatively large dispersion with
respect to metallicity has been found at a level of $^3$He/H=$(1.9\pm 0.6)\times
10^{-5}$.  It is not yet understood, however, whether $^3$He has increased or
decreased through the course of stellar and galactic evolution~\cite{Chiappini:2002hd,Vangioni-Flam:2002sa}.  Whatever the case, however, the lack of observational evidence for  the predicted galactic abundance gradient \cite{Romano03} supports the notion  that the cosmic average $^3$He abundance has not diminished from that
produced in BBN by more than a factor of $2$ due to processing in stars.  This is contrary to the case of deuterium for which  the observations and theoretical predictions  are consistent with a net decrease since the time of the BBN epoch.  Moreover, there are results \cite{Eggleton06} from 3D modeling of the region above the core convective zone for intermediate-mass giants which suggest that in net,  $^3$He is neither produced nor destroyed during  stellar burning.    Fortunately, one can avoid the ambiguity in galactic $^3$He production by making use of the fact that the sum of (D + $^3$He)/H is largely unaffected by stellar processing.  This leads to  a best estimate \cite{BBNrev} of $^3$He/H $= (0.7 \pm 0.5) \times 10^{-5}$ which implies a  reasonable  $2 \sigma$ upper
limit  of
\begin{equation}
 ^3{\rm He}/{\rm H}< 1.7 \times 10^{-5}~~.
 \label{eq22s}
\end{equation}
and a lower limit consistent with zero.

\subsubsection{$^4$He}
The primordial $^4$He abundance, $Y_p$ is best determined from HII
regions in metal-poor irregular galaxies extrapolated to zero metallicity.  A
primordial helium abundance of $ Y_p=0.247 \pm 0.002_{\rm stat} ~ \pm~ 0.004_{\rm syst}$  was deduced in Ref.~\cite{BBNrev} based upon an analysis~\cite{Peimbert:2007vm} that included new observations
and photoionization models for metal-poor extragalactic HII regions, along with  new atomic physics computations of the recombination coefficients for HeI and 
the collisional excitation of the HI Balmer lines.  However, the extraction of the final helium abundance has been fraught with uncertainties due to correlations among errors in the neutral hydrogen determination and the inferred helium abundance.  In \cite{Aver10} it was demonstrated that updated emissivities and the neutral hydrogen corrections
generally increase the inferred abundance, while the correlated uncertainties increase the uncertainty in the final extracted helium abundance.  Therefore, we adopt the value and uncertainty from \cite{Aver10} of $Y_p = 0.2561 \pm 
0.0108$, which is in general  agreement with the predicted  value from standard BBN when $\eta$ is fixed from the WMAP analysis \cite{WMAP}.  Hence, for the $Y_p$ constraint we adopt 
\begin{equation}
0.245 < Y_p < 0.267~~.
\label{eq23s}
\end{equation}

\subsubsection{$^7$Li}
 The primordial abundance of $^7$Li is best determined from old metal-poor halo stars at temperatures corresponding to the Spite plateau (see \cite{BBNrev} and references therein).  There is, however, an uncertainty in this determination due to the fact that the surface lithium in these stars may have experienced gradual depletion due to mixing with the higher-temperature stellar interiors over the stellar lifetime.  On the other hand, there are limits on the amount of such depletion that could occur since most lithium destruction mechanisms would imply a larger dispersion in abundances determined from stars of different masses, rotation rates, magnetic fields, etc., than that currently observed.   In view of this uncertainty a reasonable  upper limit on the $^7$Li abundance has been taken \cite{BBNrev}  to be $6.15\times 10^{-10}$
which is based upon allowing for  a possible depletion of up to a factor of $\sim 5$ down to the present observationally determined value of
$^7$Li/H$=(1.23^{+0.68}_{-0.32})\times 10^{-10}$ (95 \% confidence limit)~\cite{Ryan:1999vr}.  A lower
limit can be taken from the $ 2 \sigma$ observational uncertainty in the presently
observed value.  We here adopt this constraint   \cite{BBNrev} which implies  a $^7$Li/H range of
\begin{equation}
0.91\times 10^{-10} < {\rm ^7Li/H} < 6.15\times 10^{-10}~~.
\label{eq25s}
\end{equation}

\section{ Results}
 
\begin{figure}
\begin{center}
\includegraphics[width=8.0cm,clip]{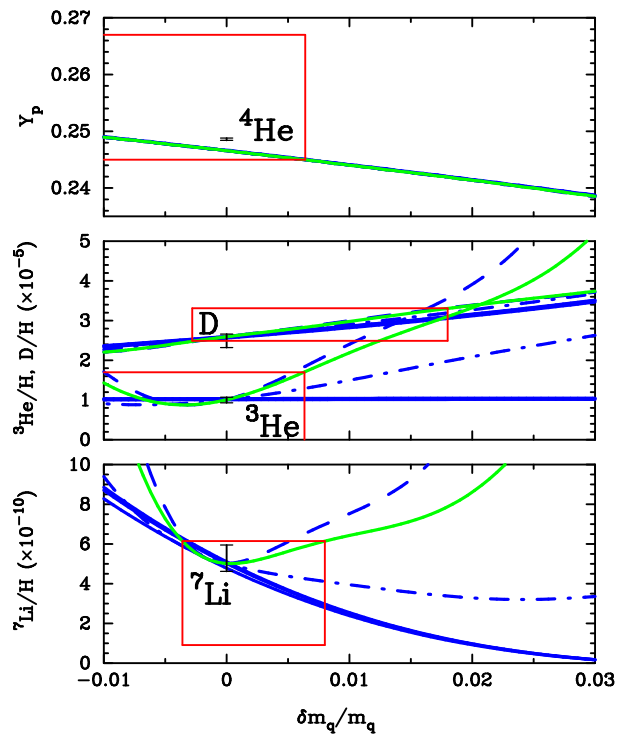}
\end{center}
\caption{Calculated light-element abundances as a function of variations
 in the quark mass  $\delta m_q/m_q$ for a fixed  $\eta = 6.23 \times
 10^{-10}$ from the  WMAP 7 year analysis \cite{WMAP}. The blue solid lines on this
 figure are for the case of no shifts in the resonance energies as in
 \cite{Berengut10}, but for the three resonant reaction rate evaluations \cite{Cyburt:2004cq,Descouvemont:2004cw,Cyburt10}.   
 The dashed line corresponds to the resonances being
 shifted the same energy as the ground state.  The dot-dashed line corresponds
 to an averaged value of the resonance sensitivity in the forward
 direction.  The solid green curve is new to the present study.  It
 derives from considering the reverse reaction for the determining the
 variation of the resonance energy with $\delta m_q/m_q$.  
 The red boxes show the  allowed parameter regions for the case of the reverse
 reaction determined using our adopted observational constraints.
 Theoretical uncertainties in standard BBN \cite{Cyburt08} are
 shown by error bars at $\delta m_q/m_q=0$ as a guide.}\label{fig:2}
\end{figure}

In  Fig. \ref{fig:2} we show primordial abundances as a function of
 variations in the quark mass $\delta m_q/m_q$ for a fixed  value of $\eta =
6.23 \times 10^{-10}$ deduced from the  WMAP 7 year data, i.e.~$\Omega_b h^2=0.02258^{+0.00057}_{-0.00056}$ for model $\Lambda$CDM+SZ+lens. 
Our calculation of standard BBN ($\delta m_q/m_q=0$) predicts the following primordial abundances: D/H=$2.593\times 10^{-5}$, $^3$He/H=$1.007\times 10^{-5}$, $Y_p=0.2466$, $^6$Li/H=$1.190\times10^{-14}$, and $^7$Li/H=$5.017\times 10^{-10}$.  We note that our calculated standard BBN abundances differ slightly from those calculated previously
in \cite{Cyburt08}.  Values and corresponding error bars from \cite{Cyburt08} for respective light nuclei at $\delta m_q/m_q=0$ are shown as points on Fig. \ref{fig:2}.
The uncertainties in the standard BBN abundances were estimated in \cite{Cyburt08} from a
Monte Carlo simulation incorporating the uncertainties in  nuclear reaction rates.
Our standard BBN $^4$He abundance is smaller than that obtained in \cite{Cyburt08} by an amount which exceeds the estimated
BBN uncertainty.  This difference mainly derives from
our adoption of the shorter  neutron lifetime from \cite{Serebrov10}.  

The blue solid
lines on Fig. \ref{fig:2} are for the case of no shifts in the resonance
energies as was also considered in \cite{Berengut10}.  For these cases we have $K_{^5{\rm Li}^*}=-1.54$ and 
$K_{^5{\rm He}^*}=-1.44$ with $\delta E_r^{(d,N)}=0$.  
All three  resonant reaction rate evaluations of  \cite{Cyburt:2004cq,Descouvemont:2004cw,Cyburt10} were utilized.  As noted above, however, the results for the different rate evaluations \cite{Cyburt:2004cq,Descouvemont:2004cw,Cyburt10} are nearly indistinguishable from those  obtained using  the earlier rate deduced in  \cite{Cyburt:2004cq}.  Hence, previous studies were  justified in using the rates from \cite{Cyburt:2004cq}.  The changes due to implementing the different rates are smaller than the drawn lines and much smaller than the uncertainties due to the errors in the BBN reaction rates as noted on Fig. \ref{fig:2}.  

The dashed lines on Fig. \ref{fig:2} show results when the resonance energies are shifted by the same amount as the ground-state binding energy shift so that the excitation energy in the compound nucleus remains fixed.  In these
cases we have $K_{^5{\rm Li}^*}=-3.35$ and 
$K_{^5{\rm He}^*}=-3.19$ and $\delta E_{^5A^*}= \delta E_{^5A_{\rm
g.s.}}$, where $^5A_{\rm g.s.}$ denotes the ground state of the $A$=5 nuclei.  

The dot-dashed lines correspond to an  average value of the
resonance sensitivity based upon resonance parameters for the forward direction.  For these curves we have
$K_{^5{\rm Li}^*}=-2.29$ and $K_{^5{\rm He}^*}=-2.21$.  The solid green curves are new to
the present study.  They derive from considering the reverse reaction rates for
 determining the variation of the resonance energy with quark mass.  For this case we have
$K_{^5{\rm Li}^*}=-3.131$, $K_{^5{\rm He}^*}=-2.867$, and the constraint:
$\delta E_r^{(N,d)}=\delta E_{^5A^*} - \delta E_{^4{\rm He}} =0$ implies
$K_{^5A^*}=K_{^4{\rm He}}\times (E_{^4{\rm He}}/E_{^5A^*})=
-1.08\times (-28.30)/E_{^5A^*}$.  We note that all lines for $^4$He are indistinguishable from each other, and that the dashed line for D is indistinguishable from the green line.

 It is straightforward to understand  the shape of calculated abundances vs~$\delta m_q/m_q$.  As  $\delta m_q/m_q$ increases, $\delta Q$ for the  $^1$H($n,\gamma$)$^2$H reaction
decreases.  The neutron processing in BBN is then delayed.  This leads to somewhat
inefficient $^4$He production and a smaller $^4$He abundance.  The
neutron abundance is then higher, so that the ultimate  D
abundance is higher.  

Thus, the $^4$He abundance decreases and the D abundance increases as $\delta m_q/m_q$ increases primarily as a result
 of a decrease in the $^1$H($n,\gamma$)$^2$H rate. 
The D abundance is, however, also slightly  affected in the same direction by the $^3$He($d,p$)$^4$He rate.  
The $^3$He abundance increases as
$\vert \delta m_q/m_q\vert $ increases.  Changes in the  resonance energies in the reactions
$^3$H($d,n$)$^4$He and $^3$He($d,p$)$^4$He lead to smaller rates as
 $\vert \delta m_q/m_q \vert$ increases.  This is because  the energy levels of the resonant states, i.e., $^5$He$^*$
and $^5$Li$^*$, are farther removed from those of the energy of the  initial entrance channels \cite{Berengut10}.  Thus, as
$|\delta m_q/m_q|$ increases, the abundances of  $^3$H and $^3$He
increase.  The abundances of $^7$Li and $^7$Be therefore also  increase since
the yields of $^7$Li [via $^4$He($^3${\rm H}$,\gamma$)] and $^7$Be [via $^4$He($^3$He$,\gamma$)]
 follow the production  of $A= 3$ nuclei.

From Fig. \ref{fig:2} it is also apparent that the revised constraints on $^7$Li
and $^4$He adopted here do not confirm a concordance best fit for $\delta
m_q/m_q = 0.016 \pm 0.005$ as deduced in \cite{Berengut10}.  Rather, the optimum
concordance level is for much smaller values of $\delta m_q/m_q$
consistent with $\delta m_q/m_q = 0.$  The constraint on the 
$^4$He abundance [Eq.~(\ref{eq23s})] limits the quark-mass variation to an upper limit of  
\begin{equation}
\delta m_q/m_q \lesssim
0.007~~~~~(^4{\rm He}),
\end{equation}
The constraint from the D abundance  [Eq.~(\ref{eq21s})] corresponds to
\begin{equation}
-0.005 \lesssim \delta m_q/m_q \lesssim 0.026~~~~~({\rm D})~~.
\end{equation}
Note, also, that adopting the more tight constraint [Eq.~(\ref{eq21t})] from the
six systems with well resolved multiple Lyman-$\alpha$ lines \cite{Pettini08} slightly reduces  the D/H limits to $-0.007 < \delta m_q/m_q < 0.023$.

There is no  limit on $\delta m_q/m_q$ 
from the $^3$He constraint [Eq.~(\ref{eq22s})] if we allow for the possibility that there is  no corresponding resonance energy shift in the  $^3$H($d,n$)$^4$He and $^3$He($d,p$)$^4$He reactions.  On the other hand, if a  resonance energy shift with $\delta m_q/m_q$ is allowed, then the $^3$He constraint could lead to an upper limit as small as 
$\delta m_q/m_q \lesssim 0.006$.

The constraint [Eq.~(\ref{eq25s})]  from $^7$Li leads only to a lower limit if we conservatively adopt only the averaged value of the forward resonance sensitivity (dot-dashed line on Fig. \ref{fig:2}):
\begin{equation}
  -0.005 < \delta m_q/m_q~~~~~(^7{\rm Li})~~.
\end{equation}
     Combining  the above limits (except $^3$He), we deduce  conservative concordance upper and lower limits for $\delta m_q/m_q$ of 
\begin{equation}
-0.005 \lesssim \delta m_q/m_q \lesssim 0.007~~~~~({\rm conservative}).
\label{eq:limit1}
\end{equation}

If more precise dependences of the rates upon the quark mass could be determined
theoretically, a stronger limit on the quark-mass variation is
possible.  For example, if the case of a resonance shift from the reverse reaction (solid green line) is
considered we obtain the limits of 
\begin{equation}
\delta m_q/m_q < 0.007~~~~~(^4{\rm He}),
\end{equation}
\begin{equation}
-0.003 < \delta m_q/m_q < 0.018~~~~~({\rm D}),
\end{equation}
\begin{equation}
\delta m_q/m_q < 0.006~~~~~(^3{\rm He}),
\end{equation}
\begin{equation}
-0.004 < \delta m_q/m_q < 0.008~~~~~(^7{\rm Li}).
\end{equation}
The combined concordance limit, however,  is not much  different from the conservative one
[Eq.~(\ref{eq:limit1})].  This is because the quark-mass upper limit is primarily constrained by  the $^4$He abundance, while 
the lower limit is constrained from
the D (or $^7$Li) abundance.  Neither  of these limits are strongly affected by the resonance energy shifts.  Nevertheless,  including the resonance energy shifts strengthens the concordance constraints.

We note that since this paper was submitted, it has come to our attention that Bedaque, Luu, \& Platter \cite{Bedaque:2010hr} independently estimated the relation between quark masses and the binding energies of light nuclides.  They reported a constraint of $-0.01 \lesssim \delta m_q/m_q \lesssim 0.007$ consistent with our conclusions.  Their analysis, however, differs from ours in several ways.  For one, their constraint is based only upon the $^4$He abundance.  Moreover, they adopted a primordial helium constraint of $0.240 \lesssim Y_p \lesssim 0.258$ that is more stringent than the recent value [Eq. (22)] adopted in the present work.  Since no reference was given, it is difficult to assess whether this more stringent value is justified.  Another difference is that the authors did not treat variations in nuclear reaction rates except for the $^1$H($n,\gamma$)$^2$H reaction.  Also, they adopted a neutron lifetime of 885 s.  Although no reference was given, this value is consistent with the old value \cite{PDG}, and different from the value \cite{Serebrov10} adopted here based upon more recent analysis.  Although it is encouraging that they have reached a similar conclusion regarding a small value for $\delta m_q/m_q$, we believe that the results reported here are based upon a more recent and thorough analysis of abundance constraints and reaction rates. 

We note on Fig. 2 that the $^4$He and D abundances exhibit a nearly linear variation with $\delta m_q/m_q$.  Since these two abundances mainly determine the concordance region,  it is useful for future reference to present here analytic formulas for the dependence of the $^4$He and D abundances with quark mass.  For the primordial helium abundance we find 
\begin{equation}
Y_p = 0.247+2.0\times10^{-4}\left(\tau_{\rm n}-878.5~{\rm s}\right)-0.26\frac{\delta m_q}{m_q}~~,
\end{equation}
where 
the dependence of $Y_p$ on the neutron life $\tau_n$ is also included. Note that this result is independent of which parametrization for the resonance shift is employed.
For the deuterium abundance, in the case where the  resonance shift in the $^3$H($d,n$)$^4$He and $^3$He($d,p$)$^4$He reactions is obtained from the reverse reactions (solid green lines in Fig. \ref{fig:2}), we obtain
\begin{equation}
{\rm D}/{\rm H}= \left(2.59+39.2\frac{\delta m_q}{m_q}\right)\times10^{-5}~~,
\end{equation}
while using the resonant shifts from the forward reactions (blue dot-dashed line) we obtain
\begin{equation}
{\rm D}/{\rm H}= \left(2.59+34.6\frac{\delta m_q}{m_q}\right)\times10^{-5}~~.
\end{equation}
If we ignore the resonance shifts (solid blue line) we obtain
\begin{equation}
{\rm D}/{\rm H}= \left(2.60+28.3\frac{\delta m_q}{m_q}\right)\times10^{-5}~~.
\end{equation}

Finally, it is worth remarking on the abundance of $^6$Li in the present studies.  This nuclide is currently of interest as there has been some suggestion in observations  \cite{Asplund:2005yt} of metal-poor halo stars for the presence of a primordial $^6$Li abundance that is $\approx 10^3$ times the standard BBN prediction.  Indeed, $^6$Li production in  standard BBN occurs via the $^4$He($d,\gamma$)$^6$Li reaction and one might expect some consequence of a varying quark mass on its abundance.   Nevertheless, the $^6$Li abundance remains small in all of the parameter space considered here. The energy of the resonance is rather high (0.712 MeV) relative to the $^4$He+$d$ entrance channel.  Moreover, its width is small,  $\Gamma = 0.024 \pm 0.002$ MeV \cite{Tilley:2002}.  Increasing the $^6$Li abundance by a significant amount would require shifting  this resonance by $\sim 0.7$ MeV.  This  would, however,  imply  an unrealistically  large $\delta m_q/m_q$.   For example, if the $K$ value for the $^6$Li resonance is the same as that for the $^6$Li ground state, the shift in the $^4$He($d,\gamma$)$^6$Li resonance energy is  given by \cite{Dent07,Berengut10}
\begin{eqnarray}
\delta E_r&=&\left( K_{^6{\rm Li}}E_{^6{\rm Li}}   - K_{^4{\rm He}}E_{^4{\rm He}} - K_d E_d \right)\frac{\delta m_q}{m_q}\nonumber\\
  &=&9.85~{\rm MeV} \frac{\delta m_q}{m_q}~~.
\end{eqnarray}
Hence, $\delta E_r=-0.7$ MeV would correspond to $\delta m_q/m_q=-0.07$ MeV.  As can be seen in Fig. 2, such a large shift in the quark mass  would result in inconsistent  abundances for $^4$He and D. Since this large deviation seems to be already excluded, the effects of a varying quark mass on the resonant  $^4$He($d,\gamma$)$^6$Li reaction can  be safely neglected.

\section{Conclusions}\label{sec4s}
We have reinvestigated effects of a hypothetical time varying quark mass on the light elements produced during big bang nucleosynthesis.  The limits on such quark-mass variations are particularly sensitive to the adopted observational abundance constraints.  Hence, in the present study we have considered updated light-element abundances and uncertainties deduced from observations.   We also consider  updated reaction rates and  an independent parametrization of the variation in the resonance energy in the $^3$He($d,p$)$^4$He reaction from the reverse $^4$He($p,d$)$^3$He reaction.   We find that the revised primordial abundances and constraints  imply that there is no concordance best fit for $\delta m_q/m_q = 0.016 \pm 0.005$ as deduced in \cite{Berengut10}.  Rather, the optimum conservative concordance region is for much smaller values of $-0.005 \lesssim \delta m_q/m_q \lesssim 0.007$, which is  consistent with $\delta m_q/m_q = 0.$

\begin{acknowledgments}
This work was supported by Grants-in-Aid for
 Scientific Research of the JSPS (20244035), and for Scientific Research on
 Innovative Area of MEXT (20105004) and the Heiwa Nakajima Foundation. Work at the University of Notre Dame was
 supported by the U.S. Department of Energy under Nuclear Theory Grant
 No. DE-FG02-95-ER40934.  Cheoun's work was supported by the National Research Foundation of Korea (Grant No. 2011-0003188).  Kusakabe's work was supported by JSPS Grant-in-Aid under Contract No. 21.6817.
\end{acknowledgments}



\end{document}